\documentclass[letterpaper,10pt]{article} 

\usepackage{opticameet3} 

\newcommand\authormark[1]{\textsuperscript{#1}}

\usepackage{amsmath,amssymb,dsfont}
\usepackage[colorlinks=true,bookmarks=false,citecolor=blue,urlcolor=blue]{hyperref} 
\usepackage[nolist]{acronym}
\usepackage{orcidlink}
\usepackage[caption=false,font=normalsize]{subfig} 

\usepackage{caption}
\usepackage{booktabs}
\usepackage{overpic}

\acrodef{IAB}{Integrated Access and Backhaul}
\acrodef{SBS}{Small Base Station}
\acrodef{MBS}{Macro Base Station}
\acrodef{BS}{Base Station}
\acrodef{UE}{User Equipment}
\acrodef{LoS}{Line-of-Sight}
\acrodef{NLoS}{Non-Line-of-Sight}
\acrodef{SINR}{signal-to-interference-plus-noise ratio}
\acrodef{GA}{Genetic Algorithm}
\acrodef{DRL}{Deep Reinforcement Learning}
\acrodef{FSO}{free space optics}
\acrodef{CU}{cost units}

\begin{document}
\title{User-Mobility-Aware Optimization of Fiber Placement in Hybrid Fiber–IAB Networks}

\author{Piotr Lechowicz~\orcidlink{0000-0003-2555-5187}~\authormark{1, *}, Charitha Madapatha~\orcidlink{0000-0003-3364-282X}~\authormark{1}, Carlos Natalino~\orcidlink{0000-0001-7501-5547}~\authormark{1}, \\Tommy Svensson~\orcidlink{0000-0002-0521-3107}~\authormark{1}, Paolo Monti~\orcidlink{0000-0002-5636-9910}~\authormark{1}}

\address{\authormark{1}Department of Electrical Engineering, Chalmers University of Technology, Gothenburg, Sweden}

\email{\authormark{*}piotr.lechowicz@chalmers.se} 

\begin{abstract}
Metaheuristic optimization of hybrid fiber–IAB networks demonstrates that integrating user dynamics into topology design enables more adaptive and cost-efficient backhaul architectures, contributing to the development of scalable and flexible 6G network infrastructures.\\
This is the authors' version of this publication. The final published version is available at \url{https://doi.org/10.1364/OFC.2026.W1H.7}.
\end{abstract}

\vspace{10pt}
\section{Introduction}
\vspace{-3pt}

6G networks necessitate the design of dynamic and resilient access topologies that integrate multiple confluent backhaul technologies to provide ubiquitous coverage with high data rates and ultra-reliable connectivity~\cite{Raj_2024_eucnc}. 
To meet these demanding requirements, mobile networks will increasingly rely on the densification of wireless \acp{BS} achieved by deploying multiple low-power \acp{SBS} closer to end users, leveraging millimeter-wave and sub-terahertz transmission spectra to increase frequency reuse. 
However, such densification poses significant challenges for the x-haul transport network, since providing fiber-optic connectivity to every \ac{SBS} remains both economically and logistically prohibitive.

From the mobile network perspective, existing solutions to mitigate these constraints are mainly based on the concept of \ac{IAB}. 
Using in-band \ac{IAB}, an \ac{SBS} connects to a \ac{MBS} (i.e., \ac{IAB} donor) using the same frequency resources that are normally assigned to serve end users.
The donor provides wireless backhaul connectivity to one or more \ac{IAB} child nodes, enabling multi-hop connections through the access spectrum.
While this approach requires no additional infrastructure, increased reliance on \ac{IAB} reduces the spectrum available for access, degrading user coverage and quality of service.
Previous studies have shown that the performance of \ac{IAB}-based networks can be significantly improved by selectively adding fiber connections between certain base stations and their donors.
Even without connecting all base stations via fiber, properly choosing a subset of \acp{SBS} for fiber backhaul can already enhance user coverage~\cite{Madapatha:2025:JointFiberFree}, i.e., the percentage of users whose achievable data rate exceeds a predefined threshold.
However, such design decisions are typically made assuming static user distributions, whereas in practical mobile scenarios, user movement over time affects where backhaul capacity is most beneficial.

Motivated by this observation, a relevant question arises on what will be the best hybrid design combining \ac{IAB} and fiber to maximize user coverage under a fixed number of fiber-connected \acp{SBS}, while considering the dynamics introduced by user mobility.
Related optimization problems have been explored \ac{GA}~\cite{Madapatha:2021:OnTopologyOptimization}, \ac{DRL}~\cite{Zhang:2025:Optimizing6GDense}, and joint resource placement methods~\cite{Lai:2020:ResourceAllocationNode}, but most assume static user distributions yielding topologies that degrade under realistic user mobility, an aspect that must be explicitly addressed in hybrid fiber-\ac{IAB} design.

This work addresses this gap by introducing user-mobility-aware metaheuristic optimization strategy for the design of hybrid fiber-\ac{IAB} topologies. 
The proposed method explicitly integrates user mobility into the optimization process, identifying which \acp{SBS} should be fiber-connected under a fixed deployment budget to maximize end-user coverage over time.
Performance evaluation in a 3GPP urban-macro scenario with random waypoint mobility shows that the proposed mobility-aware metaheuristic achieves 22.8\% higher user coverage compared to greedy baselines and a 5.1\% improvement over mobility-unaware metaheuristic designs, confirming that explicitly accounting for user movement yields more efficient hybrid fiber-\ac{IAB} configurations.
Importantly, these benefits are achieved without increasing infrastructure cost, underscoring the cost-efficiency and practical relevance of mobility-aware optimization for future 6G backhaul design.

\vspace{-7pt}
\section{Problem Description and Proposed Solutions}
\vspace{-3pt}

We consider an urban macro environment with a fixed coverage area where \acp{MBS} and \acp{SBS} are deployed as hybrid \ac{IAB} systems to serve a set of \acp{UE} \cite{Madapatha:2025:JointFiberFree}. Fig.~\ref{fig:topology} shows an example of such a system. \Acp{MBS} are interconnected by fiber links that form a ring topology and serve as \ac{IAB} donors communicating with associated \ac{IAB} child nodes (part of \acp{SBS}). 
Fiber-backhauled \acp{SBS} are connected to the ring topology using a topology design algorithm similar to the one presented in~\cite{Madapatha:2025:JointFiberFree}.
\Acp{UE} follow a random waypoint mobility pattern with shuttle destinations across four quarters of the coverage area, representing a realistic urban movement. 
We consider millimeter-wave communications at 28 GHz with 1 GHz bandwidth characterized by the 5GCM Urban Macro (UMa) channel model \cite{Rappaport_2017_OverviewMilimiterWave}.
Blockages are determined by the \ac{LoS} probability model detailed in 3GPP TR 38.901 \cite{ETSI2020}. 
\Acp{UE} are associated with a serving \ac{BS} based on the received power, calculated as the product of the transmit power, small-scale fading modeled as a Rayleigh random variable, antenna gain, and path loss.
In case a \ac{SBS} relies on \ac{IAB}, both access and backhaul links share the same, in-band, frequency bandwidth $W$. 
This bandwidth is partitioned between the access link $W_{ac}=(1-\beta)W$, and the backhaul link $W_{bh}=\beta W$, according to the ratio parameter $\beta \in \lbrack 0, 1\rbrack$.

Let $\mathbf{D}$, $\mathbf{C}$, $\mathbf{S}$, and $\mathbf{U}$ denote the sets of \ac{IAB} donors, \ac{IAB} child nodes, fiber-backhauled \acp{SBS}, and \acp{UE}, respectively.
For any given \ac{UE} $u \in \mathbf{U}$, we denote its serving \ac{BS} as $w_u$, its received \ac{SINR} as $SINR_u$, and the \ac{SINR} of its serving \ac{IAB} child node (if applicable) as $SINR_c$.
The downlink rate $R_u$ for the \ac{UE} $u$ depends on its serving \ac{BS} type, as shown in \eqref{eq_data-rate}.
The available bandwidth is equally shared among users connected to the same \ac{BS}.
Specifically, $N_{u, d}$ is the number of \acp{UE} directly served by the associated donor $d$ in $\mathbf{D}$, and $N_{u, c}$ is the number of \acp{UE} served by the associated \ac{IAB} child $c \in \mathbf{C}$, 
$N_{c, d}$ is the total number of \acp{UE} served by all \ac{IAB} child nodes connected to the same donor as child $c$,
$N_{u, s}$ is the number of \acp{UE} served by the associated fiber-backhauled \ac{SBS} $s \in \mathbf{S}$.
This network model is similar to the one in \cite{Madapatha:2025:JointFiberFree}. 

We define a binary decision variable $x_i \in \{0, 1\}, i = 1,\ldots,N$ for each of the $N$ \acp{SBS}, where $x_i$=1 indicates that \ac{SBS} $i$ is fiber-backhauled, while $x_i$=0 indicates \ac{IAB} connectivity.
In the considered problem, we assume a fixed budget represented by the fiber penetration of $k$ fiber-backhauled \acp{SBS}. 
Let $\mathds{1}(\cdot)$ denote an indicator function.
The objective function in \eqref{eq_objective}
maximizes the coverage probability, defined as the fraction of users whose achievable downlink data rate $R_u$ exceeds a predefined threshold $\eta$.
Constraint \eqref{eq_sum} ensures fixed deployment budget (fiber penetration), allowing for exactly $k$ fiber-connected \acp{SBS}, and \eqref{eq_indices} defines $x_i$ as a binary decision variable.

\vspace{-10pt}
\noindent\begin{minipage}{.55\linewidth}
\begin{equation}
    R_u = \begin{cases}
        \frac{W_{ac}}{N_{u, d}} \log_2\left(1 + SINR_u\right), &w_u \in \mathbf{D}\\
        \min \left\{
            \begin{array}{l}
            \frac{W_{ac}}{N_{u, c}} \log_2\left(1 + SINR_u\right) \\
            \frac{W_{bh}}{N_{u, c}N_{c, d}} \log_2\left(1 + SINR_c\right)
            \end{array} 
            \right\} &w_u \in \mathbf{C} \\
        \frac{W}{N_{u, s}} \log_2\left(1 + SINR_u\right), &w_u \in \mathbf{S}
    \end{cases}
    \label{eq_data-rate}
\end{equation}
\end{minipage}%
\begin{minipage}{.45\linewidth}
\begin{align}
    \mbox{maximize: } &\frac{1}{|\mathbf{U}|} \sum_{u \in \mathbf{U}} \mathds{1}\left(R_u \geq \eta\right) \label{eq_objective}\\
    \mbox{subject to: } & \sum_{i=1}^{N} x_i = k \label{eq_sum}\\
    & x_i \in \{0, 1\} \mbox{, } i = 1,\ldots,N. \label{eq_indices}
\end{align}
\end{minipage}\newline
\vspace{-3pt}

To solve this problem, we propose 2 algorithms, namely, \textit{Greedy} (GR), and \textit{Genetic Algorithm} (GA).
The algorithms are configured as follows.
GR algorithm incrementally selects $k$ \acp{SBS} based on maximizing the number of uncovered users within a 100 m distance.
At each step, it selects a \ac{SBS} that covers the most users not yet covered by any previously selected \ac{SBS}.
GA algorithm randomly creates initial solution and iteratively optimizes for the objective function.
It uses evolutionary mechanisms with a population of 20 individuals, with tournament selection of size 3, two-point crossover (rate 0.8) and swap mutation (rate 0.2), maintaining elitism by preserving the top 2 individuals across generations.

\vspace{-7pt}
\section{Results}
\vspace{-3pt}
\begin{figure}[bt!]
\begin{minipage}[b]{.5\linewidth}
\small
\footnotesize
    \centering 
    \renewcommand{\arraystretch}{0.9}
    \begin{tabular}{l l}
    \toprule
Parameter & Value  \\ \midrule
Area Size & 1km × 1km \\
MBS count & 5 \\
SBS count ($N$) & 80 \\
Fiber SBS count ($k$) & 40 \\
Users per drop & 100 \\
Number of drops & 10 \\
Time steps & 600 \\
Carrier frequency & 28 GHz \\
Bandwidth & 1 GHz \\
Beamwidth & 60° \\
Coverage threshold ($\eta$) & 100 Mbps\\
Backhaul split ($\beta$) & 0.5 \\ \bottomrule
\end{tabular}
\captionof{table}{Simulation parameters.}
    \label{tab:simulation-parameters}
  \end{minipage}\hfill
   \begin{minipage}[b]{.5\linewidth}
     \centering
    \includegraphics[width=.94\linewidth]{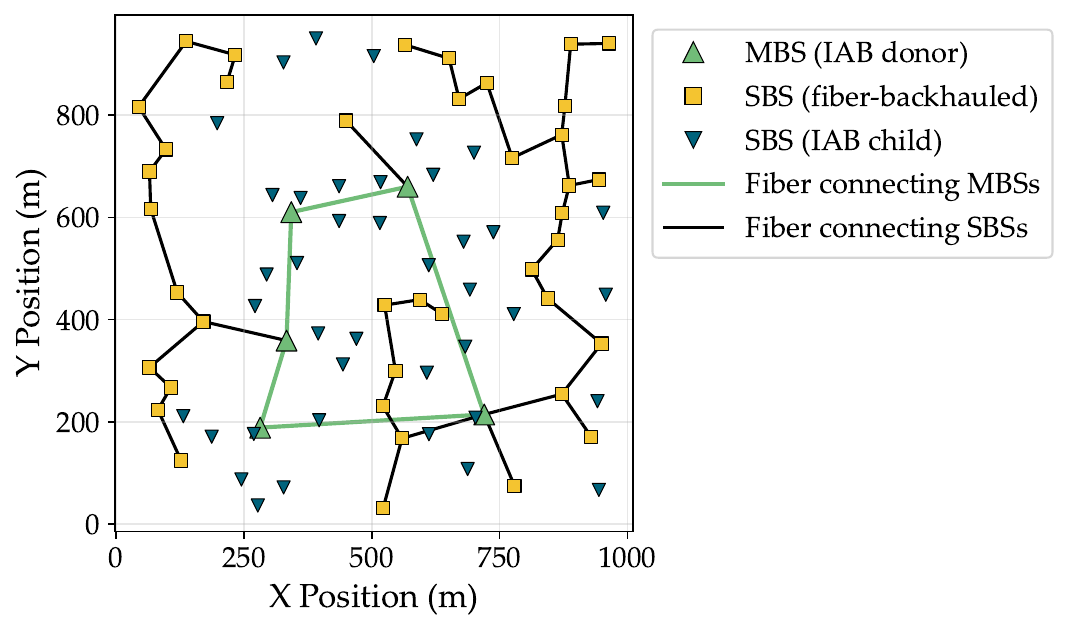}
   \captionof{figure}{Example of hybrid fiber-IAB topology.}
   \label{fig:topology}
   \end{minipage}
 \end{figure}

Table~\ref{tab:simulation-parameters} summarizes the key simulation parameters used. 
The network consists of 5 \acp{MBS} and 80 \acp{SBS}.
We consider the case where fiber penetration is set at 50\%, i.e.,  $k$=40 \acp{SBS} are connected through fiber. 
This configuration was selected based on the results presented in~\cite{Madapatha:2025:JointFiberFree}, which showed that a 50\% fiber penetration offers a favorable trade-off between deployment cost and user coverage improvement.
\Acp{MBS} and \acp{SBS} operate with 40 dBm and 24 dBm transmit power, respectively. The antenna gain follows a simplified sector antenna pattern with half-power beamwidth of 60\textdegree, 24 dBm main-lobe, and -2 dBm side-lobe gains. The data-rate threshold is $\eta$=100 Mbps. 

We evaluate the performance of the algorithms using GPU-accelerated custom simulation software written in TensorFlow.
We consider 10 randomly-generated user drops, where each drop contains 100 users with coordinates selected with a uniform distribution, enough to generate results with a confidence interval of up to 2.5\% with a 95\% confidence level.
The area is divided into four equally-sized quarters, and each user is assigned a corresponding home quarter.
Half of the users move within their home quarters and the other half shuttle between their home and another selected quarter.
The exact destination within the selected quarter is chosen uniformly with a moving speed between 1--5 m/s.
After reaching the destination, each user waits between 1--3 seconds and starts moving to the next randomly selected destination.
The cost, measured in \ac{CU}, is calculated as the sum of trenching, optical fiber, and transceivers, where the total is normalized by the cost of a 10G transceiver \cite{Madapatha:2025:JointFiberFree}.
Each algorithm runs for 1500 iterations. During the optimization process, user coverage is estimated over a short simulation of 20 seconds, with user positions updated every second. This short-term evaluation helps the algorithm quickly assess and improve candidate topologies. Once the optimization converges, the resulting topology is validated through a longer simulation of 600 seconds of network operation to obtain the final performance results.

\begin{figure}[ht]
    \begin{minipage}[b]{0.32\textwidth}
        \centering
        \begin{overpic}[width=\linewidth]{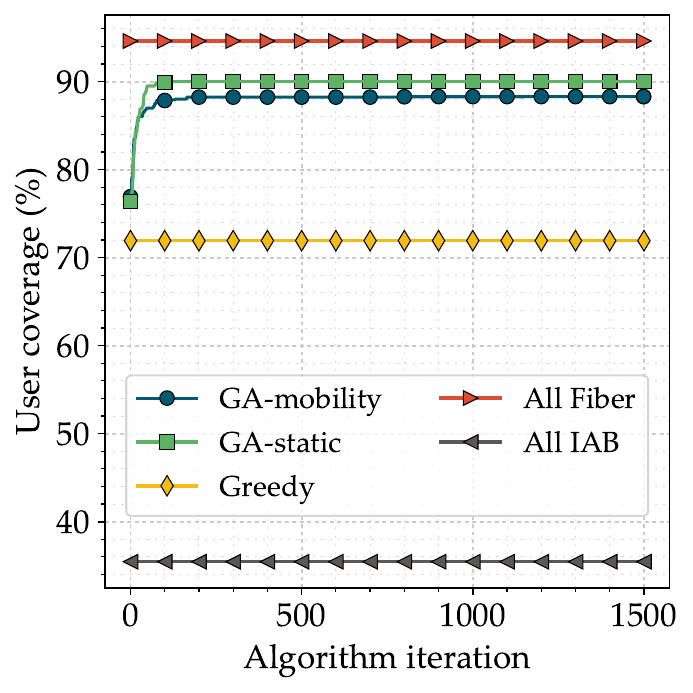} 
            \put(5, 5){a)}
        \end{overpic}
    \end{minipage}
    \hfill 
    \begin{minipage}[b]{0.32\textwidth}
        \centering
        \begin{overpic}[width=\linewidth]{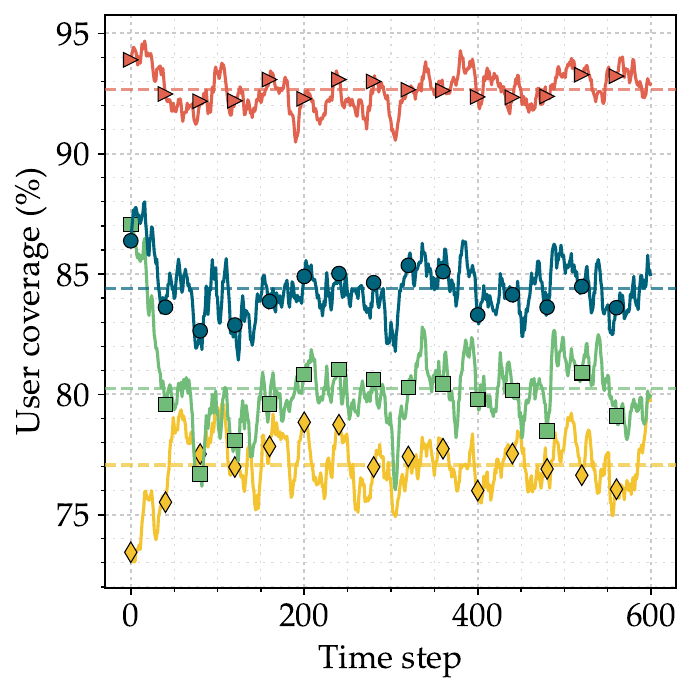} 
            \put(5, 5){b)}
        \end{overpic}
    \end{minipage}
    \hfill
    \begin{minipage}[b]{0.32\textwidth}
        \centering
        \begin{overpic}[width=\linewidth]{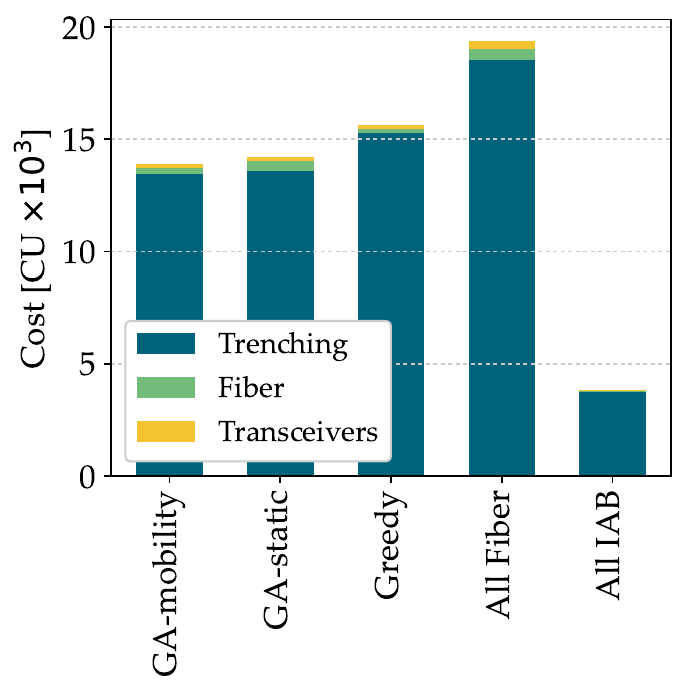} 
            \put(5, 5){c)}
        \end{overpic}
    \end{minipage}
    \caption{Hybrid fiber–IAB network design: a) User coverage versus iteration count, b) Average user coverage over time under user mobility, c) Normalized cost of the fiber infrastructure.}
    \label{fig:results}
\end{figure}

Fig.~\ref{fig:results}(a) presents the user coverage evolution as a function of algorithm iterations.
For benchmarking the following strategies are considered: \emph{All-Fiber}, where all 80 \acp{SBS} are fiber-backhauled, and \emph{All-IAB}, where all 80 \acp{SBS} operate as \ac{IAB} child nodes.
The Genetic Algorithm (GA) is executed with and without user-mobility awareness, denoted as \textit{GA-mobility} and \textit{GA-static}, respectively.
The \textit{GA-static} configuration achieves a user coverage of 90.0\%, while \textit{GA-mobility} achieves 88.3\%.
The slightly higher performance of \textit{GA-static} results from being optimized and evaluated in a purely static scenario.
Nevertheless, \textit{GA-mobility} achieves 22.8\% higher coverage compared to the Greedy algorithm.
Fig.~\ref{fig:results}(b) shows the average user coverage over time as users move across the network.
Here, the mobility-aware GA maintains a 5.1\% higher coverage than its mobility-unaware counterpart, confirming that explicitly incorporating user dynamics during the design stage enhances the network’s ability to sustain performance under user mobility.
Although moderate, a 5\% increase in user coverage can correspond to a significant number of additional users meeting their service requirements, which for operators may translate into higher resource efficiency and potential revenue gains.
Moreover, mobility-aware GA achieves a user coverage that is only 8.8\% lower than that achieved by the All-fiber case, while limiting the fiber penetration to only 50\%.
Finally, Fig.~\ref{fig:results}(c) presents the cost of the optical infrastructure, expressed in \ac{CU}.
The cost difference between designs with and without user-mobility consideration is marginal, highlighting that the improved adaptivity offered by the proposed approach is achieved without increasing deployment cost.

\section{Final remarks}

This work proposed user-mobility-aware metaheuristic optimization strategy for designing hybrid fiber-\ac{IAB} topologies. 
The method determines which \acp{SBS} should be fiber-connected to maximize end-user coverage while maintaining a fixed fiber deployment budget. 
Results show up to a 5.1\% coverage gain over mobility-unaware metaheuristic designs, and up to 22.8\% improvement compared to mobility-aware greedy baselines, while maintaining identical infrastructure cost.
Future work will extend the framework to include multi-objective optimization considering topology cost and investigate how tighter fiber deployment budgets amplify the importance of mobility awareness.

\section*{Acknowledgments}
This work has been supported by the Horizon Europe ECO-eNET project with funding from the SNS JU under grant agreement No. 101139133.

\bibliographystyle{opticaconf}
\bibliography{references.bib}

@inproceedings{Madapatha:2025:JointFiberFree,
    author = {Charitha Madapatha and Piotr Lechowicz and Carlos Natalino and Paolo Monti and Tommy Svensson},
    title = {Joint Fiber and Free Space Optical Infrastructure Planning for Hybrid Integrated Access and Backhaul Networks},
    booktitle = {Proc. of PIMRC},
    year = 2025,
pages={T4-2.2},
    eprint={2507.20367},
    archivePrefix={arXiv},
    primaryClass={cs.NI},
    url={https://arxiv.org/abs/2507.20367},
}

@ARTICLE{Madapatha:2021:OnTopologyOptimization,
  author={Madapatha, Charitha and Makki, Behrooz and Muhammad, Ajmal and Dahlman, Erik and Alouini, Mohamed-Slim and Svensson, Tommy},
  journal={IEEE Open Journal of the Communications Society}, 
  title={On Topology Optimization and Routing in Integrated Access and Backhaul Networks: A Genetic Algorithm-Based Approach}, 
  year={2021},
  volume={2},
  number={},
  pages={2273-2291},
  doi={10.1109/OJCOMS.2021.3114669}
}

@ARTICLE{Lai:2020:ResourceAllocationNode,
  author={Lai, Jiun Y. and Wu, Wu-Hsiu and Su, Yu T.},
  journal={IEEE Access}, 
  title={Resource Allocation and Node Placement in Multi-Hop Heterogeneous Integrated-Access-and-Backhaul Networks}, 
  year={2020},
  volume={8},
  number={},
  pages={122937-122958},
  doi={10.1109/ACCESS.2020.3007501}
}

@INPROCEEDINGS{Zhang:2025:Optimizing6GDense,
  author={Zhang, Jie and Chetty, Swarna and Wang, Qiao and Sun, Chenrui and Mitchell, Paul Daniel and Grace, David and Ahmadi, Hamed},
  booktitle={Proc. of IEEE INFOCOM}, 
  title={Optimizing {6G} Dense Network Deployment for the Metaverse Using Deep Reinforcement Learning}, 
  year={2025},
  doi={10.1109/INFOCOMWKSHPS65812.2025.11152919}
}

@INPROCEEDINGS{Raj_2024_eucnc,
  author={Raj, Rishu and Dass, Devika and Kaeval, Kaida and Patri, Sai and Sleiffer, Vincent and Baeuerle, Benedikt and Heni, Wolfgang and Ruffini, Marco and Browning, Colm and Naughton, Alan and Mackey, Ruth and Mackey, David and Vukovic, Boris and Smajic, Jasmin and Leuthold, Juerg and Natalino, Carlos and Wosinska, Lena and Svensson, Tommy and Kaszubowska-Anandarajah, Aleksandra and Mesogiti, Ioanna and Pryor, Simon and Tzanakaki, Anna and Nejabati, Reza and Monti, Paolo and Kilper, Dan},
  booktitle={Proc. of EuCNC}, 
  title={Towards Efficient Confluent Edge Networks}, 
  year={2024},
  volume={},
  number={},
  pages={1163-1168},
  doi={10.1109/EuCNC/6GSummit60053.2024.10597093}}

@techreport{ETSI2020,
  author    = {{ETSI}},
  title     = {{5G; Study on channel model for frequencies from 0.5 to 100 GHz (3GPP TR 38.901 version 16.1.0 Release 16)}},
  institution = {},
  number    = {TR 38 901 V16.1.0},
  year      = {2020},
  month     = nov,
  note      = {103 pp.},
  url       = {https://www.etsi.org/deliver/etsi_tr/138900_138999/138901/16.01.00_60/tr_138901v160100p.pdf}}

@article{Rappaport_2017_OverviewMilimiterWave,
    author = {Rappaport, Theodore S. and Xing, Yunchou and MacCartney, George R. and Molisch, Andreas F. and Mellios, Evangelos and Zhang, Jianhua},
    title = {Overview of Millimeter Wave Communications for Fifth-Generation ({5G}) Wireless Networks—With a Focus on Propagation Models},
    journal = {IEEE Transactions on Antennas and Propagation},
    year={2017},
    volume={65},
    number={12},
    pages={6213-6230},
    doi={10.1109/TAP.2017.2734243}
}

\end{document}